\documentclass[11pt,a4paper]{article}
\usepackage{jheppub}

\usepackage{amsmath}
\usepackage{bbm}
\usepackage{float}
\usepackage{graphicx}	
\usepackage{hyperref}
\usepackage{mathtools}

\usepackage{multirow}
\usepackage{rotating}
\usepackage{subcaption}

\def\thetitle{Compact Perturbative Expressions For Neutrino Oscillations in Matter}
\title{\thetitle}
\hypersetup{pdftitle={\thetitle}}

\author[a,b]{Peter B.~Denton}
\author[c,d]{Hisakazu Minakata}
\author[a]{Stephen J.~Parke}

\affiliation[a]{Theoretical Physics Department, Fermi National Accelerator Laboratory, P.~O.~Box 500, Batavia, IL 60510, USA}
\affiliation[b]{Physics \& Astronomy Department, Vanderbilt University, PMB 401807, 2301 Vanderbilt Place, Nashville, TN 37235, USA}
\affiliation[c]{Instituto de F\'{\i}sica, Universidade de S\~ao Paulo, C.\ P.\ 66.318, 05315-970 S\~ao Paulo, Brazil}
\affiliation[d]{Department of Physics, Yachay Tech University, San Miguel de Urcuqu\'i, 100119 Ecuador} 

\emailAdd{peterbd1@gmail.com}
\emailAdd{hminakata@yachaytech.edu.ec}
\emailAdd{parke@fnal.gov}

\preprint{FERMILAB-PUB-16-126-T}

\abstract{
We further develop and extend a recent perturbative framework for neutrino oscillations in uniform matter density so that the
resulting oscillation probabilities are accurate for the complete matter potential versus baseline divided by neutrino energy plane.
This extension also gives the {\it exact} oscillation probabilities in vacuum for all values of baseline divided by neutrino energy.
The expansion parameter used is related to the ratio of the solar to the atmospheric $\Delta m^2$ scales but with a unique choice of
the atmospheric $\Delta m^2$ such that certain first-order effects are taken into account in the zeroth-order Hamiltonian. 
Using a mixing matrix formulation, this framework has the exceptional feature that the neutrino oscillation probability in matter
has the same structure as in vacuum, to all orders in the expansion parameter. It also contains all orders in the matter potential and
$\sin\theta_{13}$.
It facilitates immediate physical interpretation of the analytic results, and makes the expressions for the neutrino oscillation
probabilities extremely compact and very accurate even at zeroth order in our perturbative
expansion. The first and second order results are also given which improve the precision by approximately two or more orders of
magnitude per perturbative order.}


\newcommand{\Dmsqren}{\Delta m^2_{ee}}
\newcommand{\U}{U_{\rm MNS}}
\newcommand{\eps}{\epsilon}

\newcommand{\ket}[1]{|#1\rangle}

\newcommand{\cc}{c_{(\phi-\theta_{13})}}
\newcommand{\s}{s_{(\phi-\theta_{13})}}
\newcommand{\Dl}[2]{\Delta\lambda_{#1#2}}
\newcommand{\e}[1]{\times10^{#1}}

\DeclareMathOperator{\diag}{diag}
\DeclareMathOperator{\sign}{sign}

\numberwithin{equation}{subsection}

\begin{document}

\maketitle

\newpage

\section{Introduction}
\label{sec:introduction}

Neutrino oscillation based on the standard three flavor scheme provides the best possible theoretical paradigm which can describe
most of the experimental results obtained in the atmospheric, solar, reactor, and the accelerator neutrino experiments. In matter, the
propogation of neutrinos is significantly modified by the Wolfenstein matter effect  \cite{Wolfenstein:1977ue}. The theoretical
derivation and understanding of the neutrino oscillation probabilities in matter have been pursued by various means. The exact
expressions of the eigenvalues, mixing angles, and the oscillation probabilities have been obtained
\cite{Barger:1980tf,Zaglauer:1988gz,Kimura:2002wd}, albeit under the assumption of uniform matter density. But, the resulting
expressions of the oscillation probabilities are way too complex to facilitate understanding of the structure of the three flavor
neutrino oscillations. For this reason, analytic approaches to the phenomena are mostly based on variety of perturbative 
frameworks. For a comprehensive treatment of neutrino oscillation in the matter, see ref.~\cite{Blennow:2013rca}.

What is the appropriate expansion parameter in such a perturbative framework? We now know that $\sin \theta_{13}$, once used as the
expansion parameter (there are an enormous number of references, see e.g., \cite{Cervera:2000kp}), is not so small, $\sin
\theta_{13} \simeq 0.15$. Moreover, expansion around $\sin \theta_{13}=0$ misses the physics of the resonance which exists at an energy
around $E \sim 10$ GeV for earth densities. Therefore, in the environments in which the matter effect is comparable to the vacuum
mixing effect, the only available small expansion parameter known to us is the ratio of the solar-scale $\Delta m^2_\odot$ to the
atmospheric-scale $\Delta m^2_\oplus$, $  \Delta m^2_\odot / \Delta m^2_\oplus \simeq 0.03$. This framework was examined in the past,
to our knowledge in refs.~\cite{Arafune:1996bt,Cervera:2000kp,Freund:2001pn,Akhmedov:2004ny}. 

Recently, two of us, see \cite{Minakata:2015gra}, presented a new perturbative framework for neutrino oscillation in matter using a
modified $  \Delta m^2_\odot / \Delta m^2_\oplus$ expansion. We identified a unique $\Delta m^2_\oplus$ that absorb certain
``first-order" terms into the ``zeroth-order'' Hamiltonian. The resulting expansion parameter,
\begin{equation*}
\epsilon   \equiv \Delta m^2_{21}/ \Dmsqren  \quad {\rm where} \quad   \Dmsqren \equiv \Delta m^2_{31} - \sin^2\theta_{12} \Delta
m^2_{21}\,,
\end{equation*}
multiplies a particularly simple perturbing Hamiltonian with zero diagonal entries.  
This re-organization of the perturbation expansion lead to simple and compact oscillation probabilities in all channels.
The $\nu_e$ disappearance channel is particularly simple, being of a pure two flavor form.

As was noted in \cite{Minakata:2015gra}, this new perturbation expansion, while valid in most of the baseline, $L$, divided by neutrino
energy, $E$, versus matter potential plane, has issues around vacuum values for the matter potential at large values of $L/E$.  These
issues are caused by the crossing of two of the eigenvalues of the new zeroth order Hamiltonian at the solar resonance.   In this
paper, we solve these issues by performing an additional rotation of the neutrino basis in matter by introducing an additional matter
mixing angle which is identical to $\theta_{12}$ in vacuum.  With this extra rotation, the new eigenvalues of the unperturbed
Hamiltonian do not cross and the perturbing Hamiltonian remains non-diagonal and is multiplied by an additional factor which is always
less than unity and is zero in vacuum.  With this additional rotation our perturbative expansion is valid in the full $L/E$ versus
matter potential plane and the zeroth order gives the exact result in vacuum.

The sectional plan of this paper is as follows: in section \ref{sec:rotations} we describe in detail the sequence of rotations of the
neutrino basis that leads us to the simple Hamiltonian that will be used in the perturbative expansion.  The zeroth order eigenvalues
and mixing matrix are given in this section.  Then, in section, \ref{sec:perturbation} we explicitly calculate the first and second
order corrections for both the eigenvalues and the mixing matrix. In section \ref{sec:oscillation probabilities}, we give compact
analytic expressions for $\nu_e$ and $\nu_\mu$ disappearance channels as well as $\nu_\mu \to \nu_e$ appearance channel at both zeroth
and first order in our perturbative expansion.  All other channels can by obtained by unitarity.
Here we discuss the precision of the perturbative treatment.
Finally, in section \ref{sec:conclusion} there is a conclusion.
A number of technical details are contained in the appendices, see \ref{sec:technical details}.
We have also published the new \verb,Nu-Pert, code used in this paper online.\footnote{See
\url{https://github.com/PeterDenton/Nu-Pert}.}

\section{Rotations of the neutrino basis and the Hamiltonian}
\label{sec:rotations}
In this section we perform a sequence of rotations on the neutrino basis and the corresponding Hamiltonian such that the following
conditions are satisfied:
\begin{itemize}
\item
The diagonal elements of the rotated Hamiltonian are excellent approximations to the eigenvalues of the exact Hamiltonian and do
not cross for any values of the matter potential.
These diagonal elements will form our $H_0$.
\item
The size of non-diagonal elements are controlled by our small parameter, $\eps'$, which vanishes in vacuum.
The non-diagonal elements will form our perturbing Hamiltonian, $H_1$.
\end{itemize}
The first two of these rotations are identical to the rotations performed in \cite{Minakata:2015gra}, while the last rotation is needed
to deal with the remaining eigenvalue crossing at the solar resonance.
With these three rotations the resulting Hamiltonian satisfies the conditions above and leads us to a rapidly converging perturbative
expansion for the oscillation probabilities that covers all of the $L/E$ versus matter potential plane.

\subsection{Overview}
Neutrino evolution in matter is governed by a Schr\"odinger like equation
\begin{equation} 
i\frac{\partial}{\partial x}\ket{\nu}=H\ket{\nu}\,,
\label{eq:schrod}
\end{equation}
where in the flavor basis
\begin{gather}
\ket{\nu}=
\begin{pmatrix}
\nu_e\\\nu_\mu\\\nu_\tau 
\end{pmatrix}\,,\\
H=\frac{1}{2E}\left[\U\diag(0,\Delta m^2_{21},\Delta m^2_{31})\U^\dagger+\diag(a(x),0,0)\right]\,.
\end{gather}
$\U$ is the lepton mixing matrix in vacuum, given by\\
$\U\equiv U_{23}(\theta_{23},\delta)U_{13}(\theta_{13})U_{12}(\theta_{12})$ with\footnote{The PDG form of $\U$ is obtained from our
$\U$ by multiplying the 3rd row by $e^{i\delta}$ and the 3rd column by $e^{-i\delta}$ i.e.~by rephasing $\nu_\tau$ and $\nu_3$.
The shorthand notation $c_\theta = \cos \theta$ and $s_\theta = \sin \theta$ is used throughout this paper.}
\begin{equation}
\begin{gathered}
U_{12}(\psi) \equiv
\begin{pmatrix}
c_\psi&s_\psi\\
-s_\psi&c_\psi\\
&&1
\end{pmatrix}\, ,\quad
U_{13}(\phi) \equiv
\begin{pmatrix}
c_{\phi}&&s_{\phi}\\
&1\\
-s_{\phi}&&c_{\phi}
\end{pmatrix}\,,\\
U_{23}(\theta_{23}, \delta)\equiv
\begin{pmatrix}
1\\
&c_{23}&s_{23}e^{i\delta}\\
&-s_{23}e^{-i\delta}&c_{23}
\end{pmatrix} \, ,
\end{gathered}
\label{eq:MNS}
\end{equation}
and the matter potential, assumed to be constant, is given by
\begin{equation}
a\equiv2\sqrt2G_FN_eE\approx1.52\e{-4}\left(\frac{Y_e\rho}{\rm g\cdot{\rm cm}^{-3}}\right)\left(\frac E{\rm GeV}\right){\rm eV}^2\, .
\end{equation}

We will perform a sequence of rotations on the flavor basis by multiplying the left and right hand side of eq.~\ref{eq:schrod} by an
appropriate unitary matrix, $U^\dagger$ and inserting unity ($UU^\dagger$) between $H$ and $\ket{\nu}$. These rotations are
chosen such that the final resulting Hamiltonian satisfies the following properties: the diagonal elements are an excellent
approximations to the exact eigenvalues and the size of off-diagonal elements are controlled by a small parameter (ratio of the $\Delta
m^2$'s) and are identically zero in vacuum.

The sequence of rotations applied to the eigenstates is performed in the following order
\begin{equation}
\begin{aligned}
\ket{\nu}\to\ket{\tilde{\nu}}={}&U^\dagger_{23}(\theta_{23}, \delta)\ket{\nu}\\
\to{}&\ket{\hat{\nu}}=U^\dagger_{13}(\phi)U^\dagger_{23}(\theta_{23},\delta)\ket{\nu}\\
&\to\ket{\check{\nu}}=U^\dagger_{12}(\psi)U^\dagger_{13}(\phi)U^\dagger_{23}(\theta_{23},\delta)\ket{\nu}\,,
\end{aligned}
\end{equation}
with the corresponding Hamiltonians
\begin{equation}
\begin{aligned}
H\to\tilde{H}={}&U^\dagger_{23}(\theta_{23},\delta)~H~U_{23}(\theta_{23},\delta)\\
\to{}&\hat{H}=U^\dagger_{13}(\phi)U^\dagger_{23}(\theta_{23},\delta)~H~U_{23}(\theta_{23},\delta) U_{13}(\phi)\\
&\to\check{H}=U^\dagger_{12}(\psi)U^\dagger_{13}(\phi)U^\dagger_{23}(\theta_{23},\delta)~H~U_{23}(\theta_{23},\delta)U_{13}
(\phi)U_{12}(\psi)\,.
\end{aligned}
\end{equation}

The first rotation undoes the $\theta_{23}-\delta$ rotation, whereas the $\phi$ followed by $\psi$ rotations are matter analogues to
the vacuum $\theta_{13}$ and $\theta_{12}$ rotations, respectively. In vacuum, the final Schr\"odinger equation is just the trivial
mass eigenstate evolution equation.

\subsection{\texorpdfstring{$U_{23}(\theta_{23},\delta)$}{U23(theta23,delta)} rotation}
\label{ssec:U23}
After the $U_{23}(\theta_{23}, \delta)$ rotation, the neutrino basis is 
\begin{equation}
\ket{\tilde{\nu}}=U^\dagger_{23}(\theta_{23},\delta)\ket{\nu}\,,
\end{equation}
and the Hamiltonian is given by 
\begin{gather}
\begin{aligned}
\tilde{H}={}&U^\dagger_{23}(\theta_{23},\delta)~H~U_{23}(\theta_{23},\delta)\\
={}&\frac{1}{2E}\left[U_{13}(\theta_{13})U_{12}(\theta_{12})\diag(0,\Delta m^2_{21},\Delta m^2_{31})U^\dagger_{12}(\theta_{12})
U^\dagger_{13}(\theta_{13})\right.\\
&\left.\vphantom{U^\dagger_{1}}+\diag(a,0,0)\right]\,.
\end{aligned}
\end{gather}
As was shown in  \cite{Minakata:2015gra},  the  Hamiltonian, $\tilde{H}$,  is most simple written in terms of a renormalized
atmospheric $\Delta m^2$,
\begin{equation}
\Dmsqren\equiv\Delta m^2_{31}-s_{12}^2\Delta m^2_{21}\,,
\end{equation}
as defined in \cite{Nunokawa:2005nx,Parke:2016joa}, and the ratio of the $\Delta m^2$'s
\begin{equation}
\eps\equiv\Delta m^2_{21}/\Dmsqren \, .
\label{eq:eps}
\end{equation}
In terms of the $|a| \to \infty $ eigenvalues
\begin{gather}
\begin{aligned}
\lambda_a &=a+(s^2_{13}+\eps s^2_{12})\Dmsqren\,, \\
\lambda_b &=\eps c^2_{12}\Dmsqren\,, \\
\lambda_c &=(c^2_{13}+\eps s^2_{12})\Dmsqren\,,
\label{eq:lambda abc}
\end{aligned}
\end{gather}
the exact Hamiltonian is simple given by\footnote{One can use $\tilde{H}$ to do a perturbative expansion, such that it is simple to
recover the $\nu_\mu \to \nu_e$ appearance probability of  Cervera et al., \cite{Cervera:2000kp} at first order.}
\begin{equation}
\begin{gathered}
\tilde H=\frac{1}{2E}
\begin{pmatrix}
\lambda_a && s_{13}c_{13}\Dmsqren\\
& \lambda_b & \\
s_{13}c_{13}\Dmsqren & &\lambda_c
\end{pmatrix}
+\eps s_{12}c_{12}\frac{\Dmsqren}{2E}
\begin{pmatrix}
&c_{13}\\
c_{13}&&-s_{13}\\
&-s_{13}
\end{pmatrix}\,.
\end{gathered}
\end{equation}
Note that $\tilde{H}$ is real and does not depend on $\theta_{23}$ or $\delta$.

\subsection{\texorpdfstring{$U_{13}(\phi)$}{U13(phi)} rotation}
\label{ssec:U13}
Since $ s_{13} \sim {\mathcal O}(\sqrt{\eps})$, it is natural to diagonalize the (1-3) sector next, using $U_{13}(\phi)$, again see
\cite{Minakata:2015gra}.
After this rotation the neutrino basis is 
\begin{equation}
\ket{\hat{\nu}}=U^\dagger_{13}(\phi)\ket{\tilde{\nu}}=U^\dagger_{13}(\phi)U^\dagger_{23}(\theta_{23},\delta)\ket{\nu}\,,
\end{equation}
and the Hamiltonian is given by 
\begin{gather}
\begin{aligned}
\hat{H}&=U^\dagger_{13}(\phi)~\tilde{H}~U_{13}(\phi)\\ 
&=\frac{1}{2E}
\begin{pmatrix}
\lambda_-\\
&\lambda_0\\
&&\lambda_+
\end{pmatrix}\,
+\eps c_{12}s_{12}\frac{\Dmsqren}{2E}
\begin{pmatrix}
&\cc\\
\cc&&\s\\
&\s
\end{pmatrix}\,.
\label{eq:hatH}
\end{aligned}
\end{gather}
where
\begin{gather}
\begin{aligned}
\lambda_\mp&=\frac12\left[(\lambda_a+\lambda_c)\mp\sign(\Dmsqren)\sqrt{(\lambda_c-\lambda_a)^2+4(s_{13}c_{13}\Dmsqren)^2}\right]\,,\\
\lambda_0&=\lambda_b=\eps c_{12}^2\Dmsqren\,,
\end{aligned}
\end{gather}
which is identical to eq.~3.1 of \cite{Minakata:2015gra}.

The angle, $\phi$, that achieves this diagonalization of the (1-3) sub-matrix (see appendix \ref{ssec:general diagonalization}),
satisfies
\begin{equation}
\lambda_a=c^2_\phi \lambda_- + s^2_\phi \lambda_+\,, \quad 
\lambda_c=s^2_\phi \lambda_- + c^2_\phi \lambda_+\,, \quad {\rm and} \quad
s_\phi c_\phi = \frac{s_{13}c_{13}\Dmsqren}{\lambda_+- \lambda_-}\,,
\label{eq:s2phi}
\end{equation}
from which it is easy to derive
\begin{gather}
c^2_\phi-s^2_\phi=\frac{\lambda_c- \lambda_a}{\lambda_+- \lambda_-}\,,\\
s_\phi=\sqrt{\frac{\lambda_+-\lambda_c}{\lambda_+-\lambda_-}}\,,\qquad
c_\phi=\sqrt{\frac{\lambda_c-\lambda_-}{\lambda_+-\lambda_-}}\,.
\label{eq:c2phi}
\end{gather}
The Hamiltonian given in eq.~\ref{eq:hatH} was used to derive simple, compact and accurate oscillation probabilities for a wide range
of the $L/E$ versus $\rho E$ plane, see \cite{Minakata:2015gra}.
However, as was noted in that paper, there is a region of this plane for which a perturbation theory based on $\hat{H}$ is
insufficient to describe the physics accurately.
This region is small $\rho E$ and large $L/E$ given by
\begin{equation}
|a| < \frac{1}{3} \Dmsqren \quad {\rm and} \quad L/E >\frac{4\pi}{\Dmsqren}\,.
\end{equation}
To address this region of the $L/E$ versus $\rho E$ plane, we perform one further rotation on the Hamiltonian.
This rotation removes the degeneracy of the zeroth order eigenvalues at the solar resonance when $\lambda_- = \lambda_0$.
This is performed in the next subsection.

\subsection{\texorpdfstring{$U_{12}(\psi)$}{U12(psi)} rotation}
\label{ssec:U12}
Since $\lambda_-$ and $\lambda_0$ cross at the solar resonance, $a\approx\eps \Dmsqren \cos 2 \theta_{12}/\cos^2 \theta_{13}$, to
describe the physics near this degeneracy we need to diagonalize the (1-2) submatrix of $\hat{H}$, using $U_{12}(\psi)$.
The new neutrino basis is
\begin{equation}
\ket{\check{\nu}} = U^\dagger_{12}(\psi)\ket{\hat{\nu}} = U^\dagger_{12}(\psi) U^\dagger_{13}(\phi)U^\dagger_{23}(\theta_{23},
\delta)\ket{\nu}\,.
\end{equation}
The resulting Hamiltonian, split into a zeroth order Hamiltonian and a perturbing Hamiltonian, is given by
\begin{equation}
\check{H}=U^\dagger_{12}(\psi) ~\hat{H} ~ U_{12}(\psi) = \check{H}_0 + \check{H}_1\,,
\end{equation}
where
\begin{align}
\label{eq:Hcheck0}
\check{H}_0&=\frac{1}{2E}
\begin{pmatrix}
\lambda_1&&\\
&\lambda_2&\\
&&\lambda_3
\end{pmatrix}\, , \\
\check{H}_1&=\eps \s s_{12}c_{12} \frac{\Dmsqren}{2E}
\begin{pmatrix}
&&-s_\psi\\
&&c_\psi\\
-s_\psi&c_\psi
\end{pmatrix}\,.
\label{eq:Hcheck1}
\end{align}
The diagonal elements of the zeroth order Hamiltonian are 
\begin{gather}
\begin{aligned}
\lambda_{1,2} & = \frac12\left[(\lambda_0+\lambda_-)\mp\sqrt{(\lambda_0-\lambda_-)^2+4(\eps \cc
c_{12}s_{12}\Dmsqren)^2}\right]\, ,\\
\lambda_3 & = \lambda_+\,.
\label{eq:lambdacheck}
\end{aligned}
\end{gather}

The angle, $\psi$, that achieves this diagonalization of the (1-2) sub-matrix of $\hat{H}$ (see appendix \ref{ssec:general
diagonalization}), satisfies
\begin{equation}
\lambda_- = c^2_\psi \lambda_1 + s^2_\psi \lambda_2\,, \qquad 
\lambda_0 = s^2_\psi \lambda_1 + c^2_\psi \lambda_2\,,
\end{equation}
\begin{equation}
s_\psi c_\psi = \frac{ \eps \cc s_{12}c_{12} \Dmsqren}{\Dl21}\,,
\label{eq:s2psi}
\end{equation}
where we introduce the useful shorthand notation, 
\begin{equation}
\Dl ij\equiv\lambda_i-\lambda_j \, .
\end{equation}
It is easy to derive that\footnote{Given the definition of $\lambda_{1,2}$ in eq.~\ref{eq:lambdacheck}, the sign term in
from of $c_\psi$ is not necessary, but will become necessary when we discuss the $\lambda_{1} \leftrightarrow \lambda_{2}$ interchange
symmetry.}
\begin{gather}
\label{eq:c2psi}
c^2_\psi-s^2_\psi=\frac{\lambda_0- \lambda_-}{\Dl21} \,,\\
{\rm and} \quad s_\psi = \sqrt{ \frac{\lambda_2-\lambda_0}{\Dl21}}\,,\qquad
c _\psi= \sign(\Dl21)\sqrt{ \frac{\lambda_0-\lambda_1}{\Dl21}}\,.
\label{eq:cspsi}
\end{gather}

\begin{figure}
\centering
\includegraphics[width=0.495\textwidth]{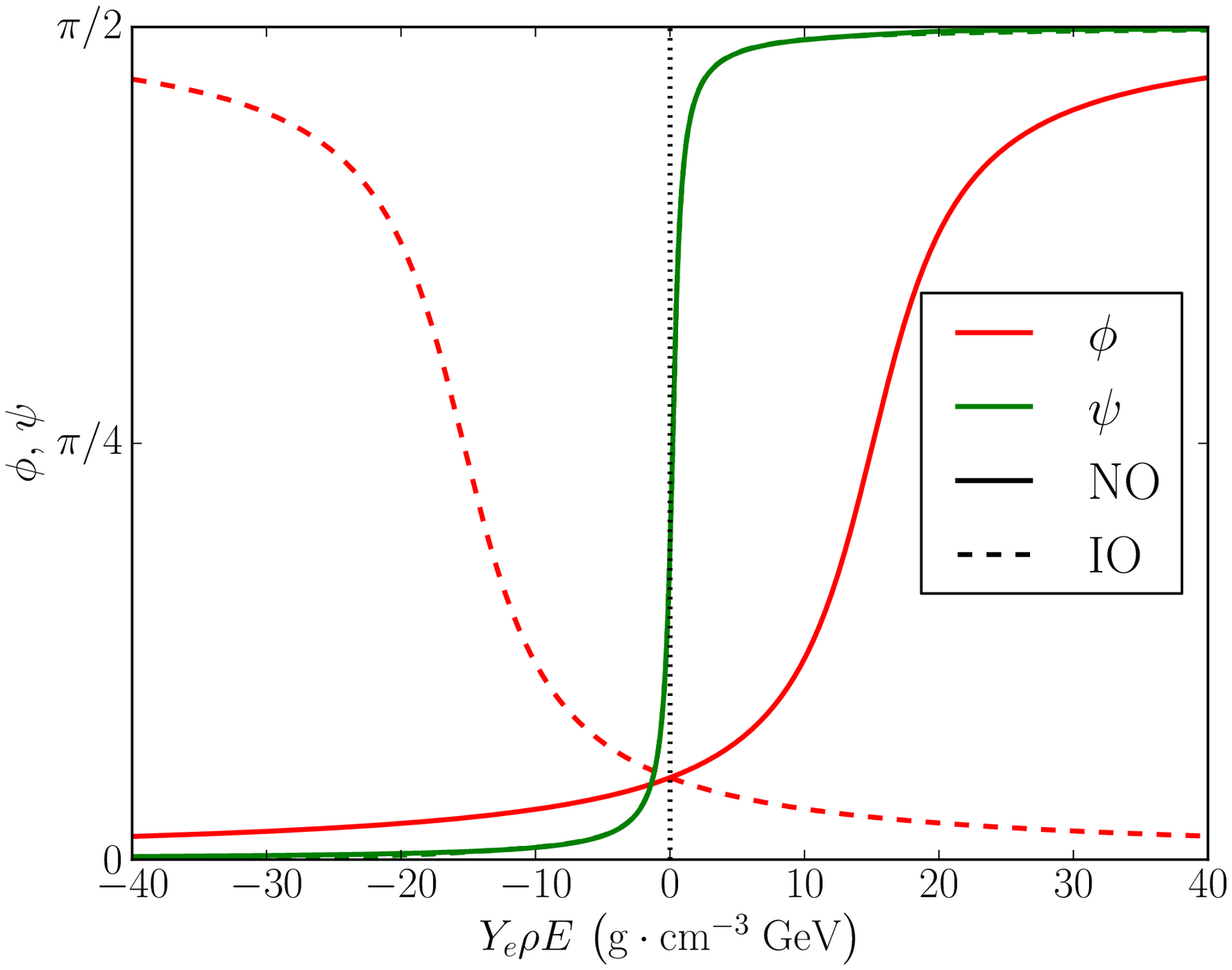}\\
\includegraphics[width=0.495\textwidth]{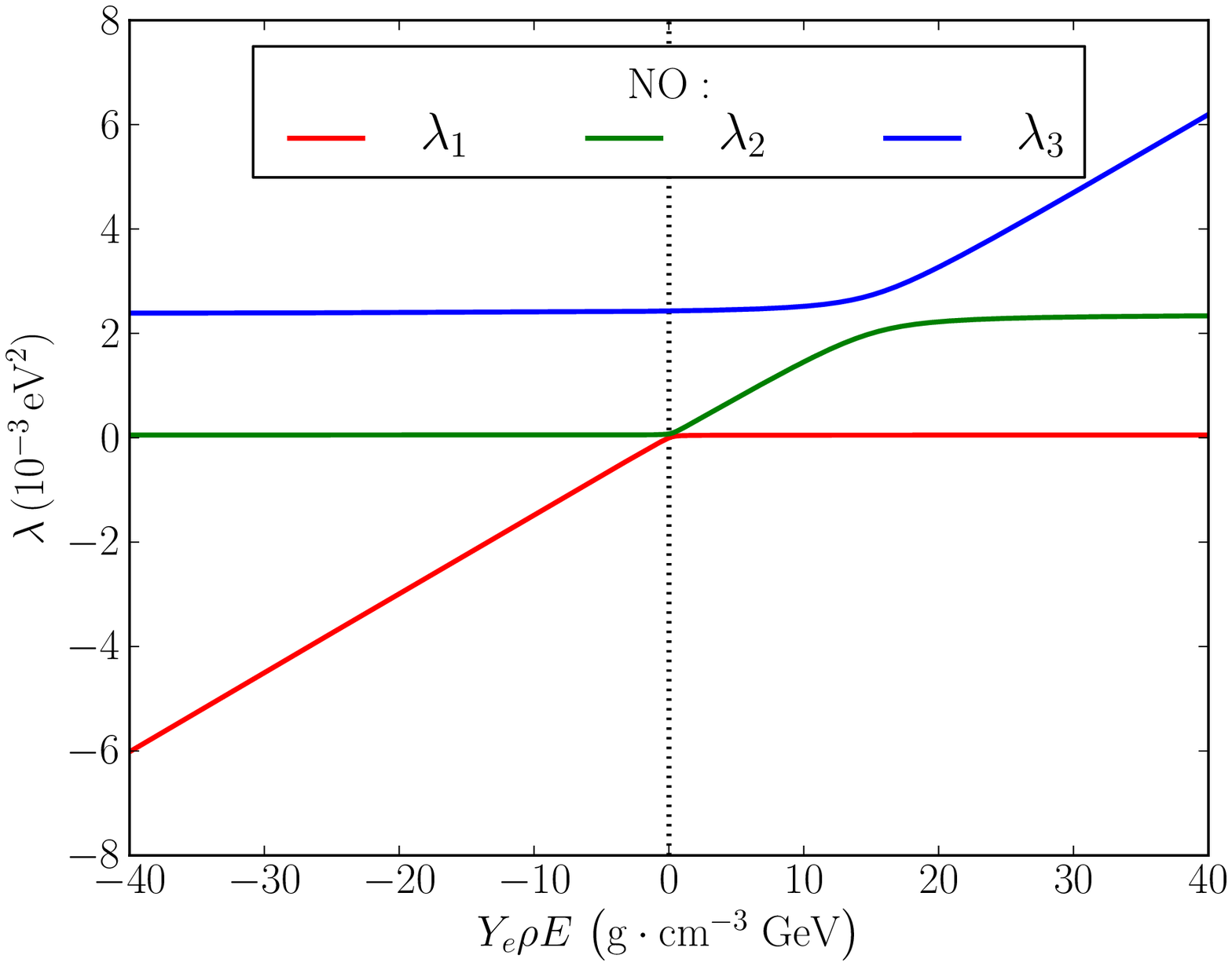}
\includegraphics[width=0.495\textwidth]{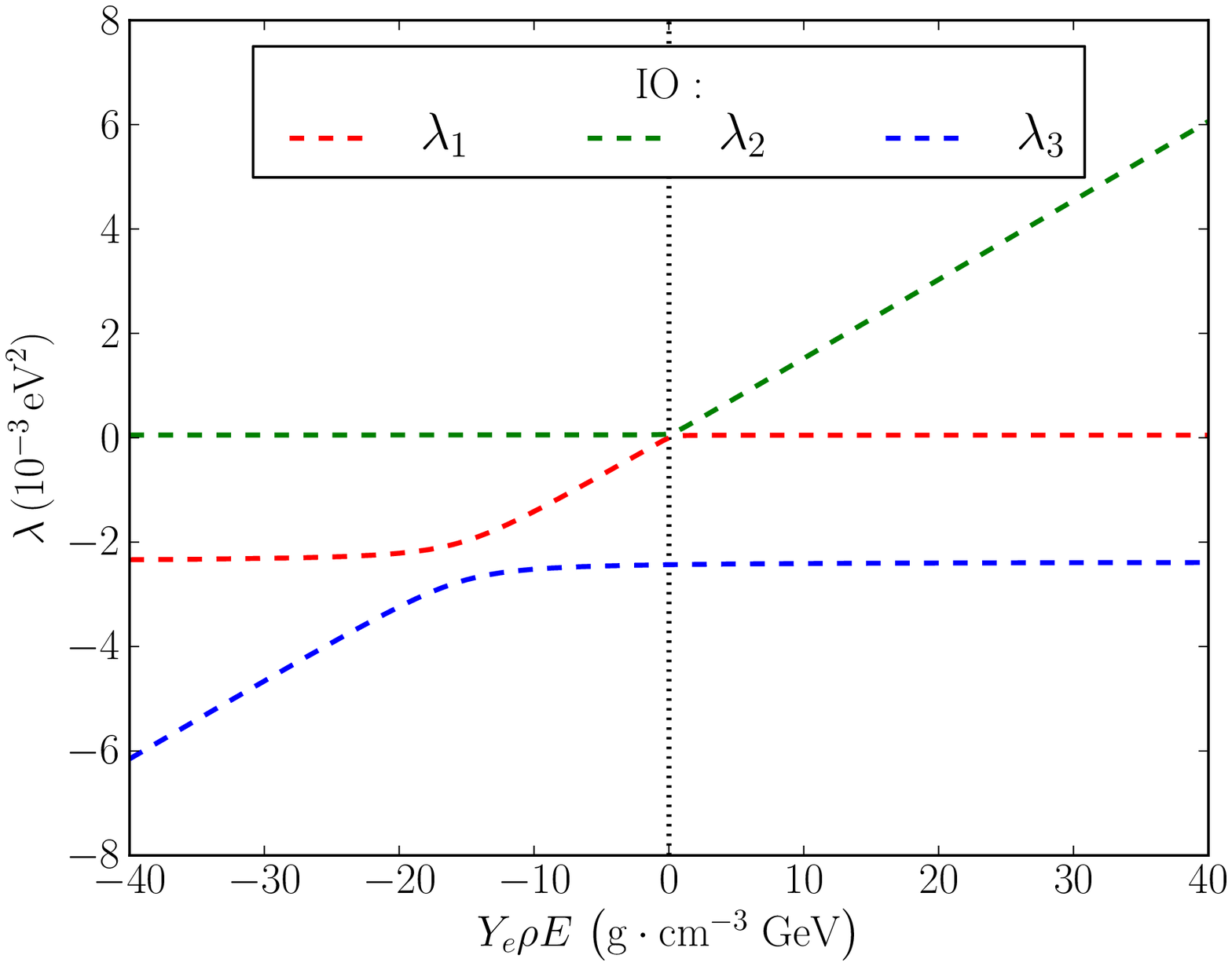}
\caption{
The upper figure shows the angles, $\phi$ and $\psi$, as a function of the matter potential for both NO and IO.
$\phi$ and $\psi$ are the mixing angles $\theta_{13}$ and $\theta_{12}$ in matter respectively.
For $\psi$, the curves for the two mass ordering are nearly identical.
The two lower figures show the eigenvalues to zeroth order, $\lambda_{1,2,3}$, in matter as a function of the matter potential
for NO and for IO.
For all our figures, $Y_e \rho E \geq 0$ is for neutrinos and \ $Y_e \rho E \leq 0$ for antineutrinos.}
\label{fig:angles+eigenvalues}
\end{figure}

Figure \ref{fig:angles+eigenvalues} shows $\phi$ and $\psi$ as functions of the matter potential as well as the eigenvalues of
$\check{H}$ for both the normal ordering (NO) and the inverted ordering (IO).
Several additional useful identities used in the calculations throughout this paper are listed in appendix \ref{ssec:useful
identities}.

\subsection{Remarks}
\begin{figure}
\centering
\includegraphics[width=0.5\textwidth]{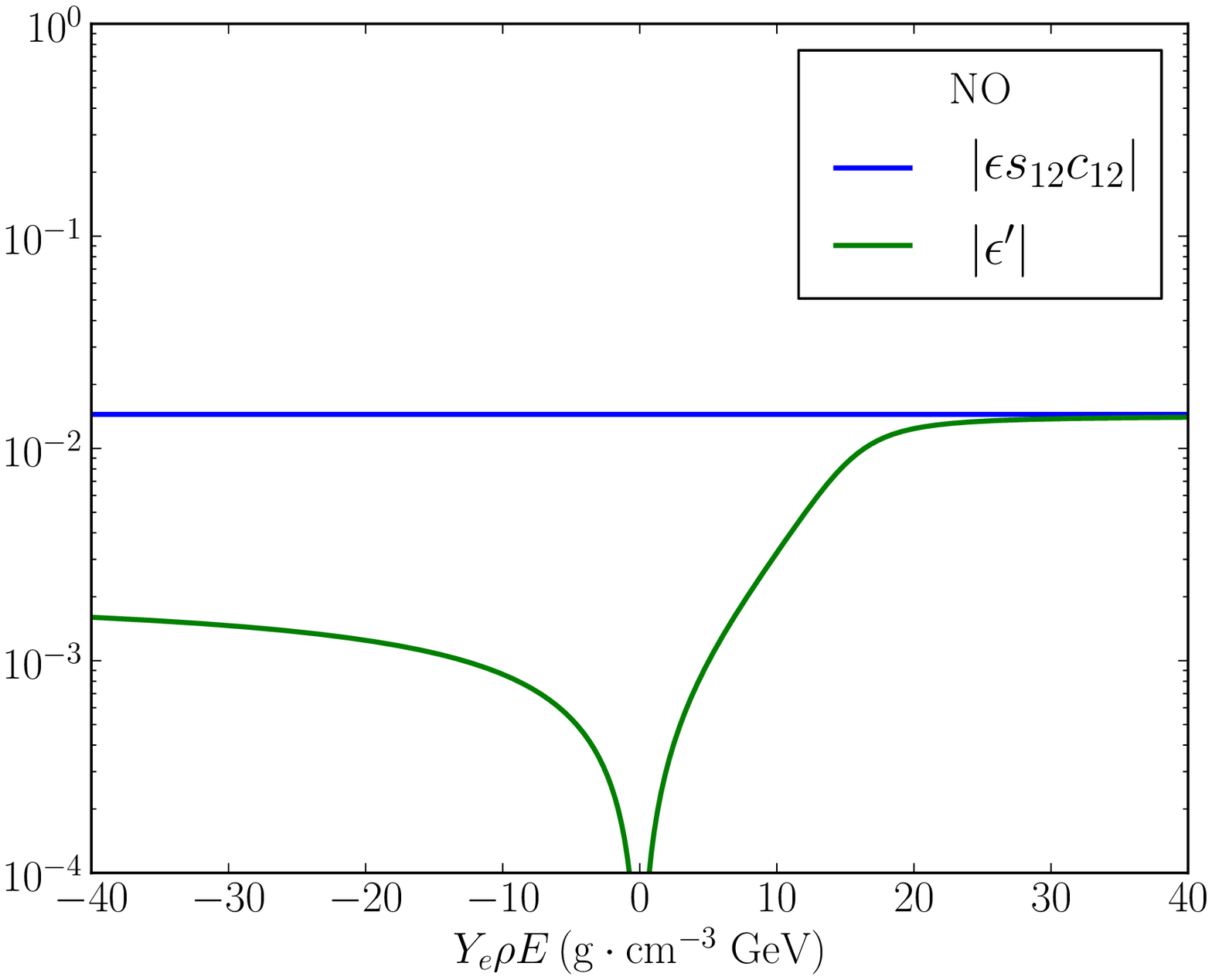}
\caption{The absolute value of the various expansion parameters as a function of the matter potential.
$\eps s_{12}c_{12}\equiv s_{12}c_{12}\Delta m^2_{21}/\Dmsqren$ is the expansion parameter from \cite{Minakata:2015gra}
and $\eps'\equiv \s  s_{12}c_{12} \Delta m^2_{21}/\Dmsqren$ is the expansion parameter of this paper, see eqs.~\ref{eq:eps}
and \ref{eq:epsp}.
The asymptotic value of $|\eps'|$ as $E\to-\infty$ is $|\eps c_{12}s_{12}s_{13}| \approx2.2\e{-3}$ and as $E\to\infty$ is $|\eps
c_{12}s_{12} c_{13}| \approx 1.4\e{-2}$.
The NO is shown here, the IO is the same with $Y_e\rho E\to-Y_e\rho E$.}
\label{fig:expansion parameter}
\end{figure}

A number of summarizing and useful comments are warranted at this point.
\begin{itemize}
\item The neutrino basis that will be used in our perturbation theory, $\ket{\check{\nu}}$ is related to the flavor basis, 
$\ket{\nu}$ by
\begin{equation}
\begin{pmatrix}
\nu_e\\ \nu_\mu\\ \nu_\tau 
\end{pmatrix}=\U^m
\begin{pmatrix}
\check{\nu}_1 \\ \check{\nu}_2\\ \check{\nu}_3 
\end{pmatrix}\,,
\end{equation}
where
\begin{equation}
\U^m \equiv U_{23}(\theta_{23}, \delta) U_{13}(\phi) U_{12}(\psi)\,.
\label{eq:UmMNS}
\end{equation}

\item
The Hamiltonian, eqs.~\ref{eq:Hcheck0} and \ref{eq:Hcheck1}, that will used as the basis for our perturbation theory is given by
\begin{equation}
\check H=(U_{\rm MNS}^{m})^{\dagger}HU_{\rm MNS}^m = \check H_0 + \check H_1\,,
\end{equation}
with the diagonal elements the zeroth order Hamiltonian and the off-diagonal elements the perturbing Hamiltonian.
While the $\lambda_{a,b,c}$ eigenvalues cross twice and the $\lambda_{-,0,+}$ eigenvalues cross once, the new $\lambda_{1,2,3}$
eigenvalues do not cross, see figure \ref{fig:angles+eigenvalues}, which allows for the perturbation theory to be well defined
everywhere.

\item
The size of the perturbing Hamiltonian, $\check{H}_1$, is controlled by the parameter
\begin{gather}
\begin{aligned}
\epsilon^\prime & \equiv \eps ~ \s ~s_{12} c_{12} \\
& = \s s_{12} c_{12} \frac{\Delta m^2_{21}}{\Dmsqren}\,,
\label{eq:epsp}
\end{aligned}
\end{gather}
which is never larger than 1.4\%.
\item
In vacuum,
\begin{equation}
\s=0\,,
\end{equation}
so that the zeroth order Hamiltonian gives the \emph{exact} result. 
Also, in the limit where $a\to -\infty$ for NO or $a\to +\infty$ for IO
$\s \to -s_{13}$
 which is of ${\mathcal O}(\sqrt{\epsilon})$. Whereas for $a\to +\infty$ for NO or $a\to -\infty$ for IO $\s \to c_{13} \sim
1$, see figure \ref{fig:expansion parameter}.

\item
Since perturbing Hamiltonian, $\check H_1$, has only non-diagonal entries the first order correction to the eigenvalues are zero.
The diagonal elements multiplied by $2E$ are, to an excellent approximation, the mass squares of the neutrinos in matter.

\item
There is a very useful interchange symmetry involving $\lambda_{1,2}$ and $\psi$.
The Hamiltonian is invariant under the pair of transformations $\lambda_1\leftrightarrow\lambda_2$ and $\psi \to\psi \pm \pi/2$.
Our expressions for $s_\psi$ and $c_\psi$, see eq.~\ref{eq:cspsi}, satisfy this interchange symmetry with the $+$ in
front of the $\pi/2$.
Since the transition probabilities always have an even number of $\psi$ trig functions, this interchange symmetry can be simply
expressed as
\begin{equation}
\lambda_1\leftrightarrow\lambda_2\,,\quad
c_\psi^2\leftrightarrow s_\psi^2\,,\quad{\rm and}\quad
c_\psi s_\psi\leftrightarrow-c_\psi s_\psi\,.
\label{eq:lambda12psi}
\end{equation}
In the rest of this paper we call this the $\lambda_{1,2}-\psi$ interchange symmetry.
\item
An antineutrino with energy $E$ is equivalent to a neutrino with energy $-E$.
\item
The values of all of the eigenvalues in vacuum and for $a\to\pm\infty$ are shown in appendix \ref{ssec:limits}.
\end{itemize}

\section{Perturbation expansion}
\label{sec:perturbation}
To calculate the neutrino oscillation probabilities at zeroth order, all that is needed is eigenvalues and mixing matrix,
\begin{equation*}
\lambda_{1,2,3}  \quad {\rm and} \quad  \U^m\,,
\end{equation*}
given by eq.~\ref{eq:lambdacheck} and eq.~\ref{eq:UmMNS} respectively.
 For higher order calculations we need not only the corrections to the eigenvalues but also the corrections to the mixing matrix.  In
this section we first given the corrections to the eigenvalues at both first and second order in our expansion parameter, $\epsilon'$.
This is followed by the corrections to the same order for the mixing matrix.  Note that all corrections to both the eigenvalues and the
mixing matrix vanish in vacuum as our expansion parameter is zero in vacuum, i.e.~the zero order oscillation probabilities are exact in
vacuum.

\subsection{Corrections to the eigenvalues}
Since the diagonal terms of $\check H_1=0$ by construction, the first order corrections to the eigenvalues are exactly zero, since
\begin{equation}
\lambda_i^{(1)}=2E(\check H_1)_{ii} = 0\,.  
\label{eq:eigenvalues to first order}
\end{equation}
The second order corrections to the eigenvalues are given by\footnote{Eq.~\ref{eq:eigenvalues to second order} explicitly shows why the
level crossing of two of the eigenvalues ($\lambda_-,\lambda_0$) causes problems for higher orders in the
perturbation theory.}
\begin{equation}
\lambda_i^{(2)}=\sum_{k\neq i}\frac{[2E(\check H_1)_{ik}]^2}{\Dl ik}\,.
\label{eq:eigenvalues to second order}
\end{equation}
Using $\check H_1$ from eq.~\ref{eq:Hcheck1}, we see that the corrections are
\begin{equation}
\begin{aligned}
\lambda_1^{(2)}&=-(\eps'\Dmsqren)^2\frac{s_\psi^2}{\Dl31}\,,\\
\lambda_2^{(2)}&=-(\eps'\Dmsqren)^2\frac{c_\psi^2}{\Dl32}\,,\\
\lambda_3^{(2)}&=(\eps'\Dmsqren)^2\left(\frac{s_\psi^2}{\Dl31}+\frac{c_\psi^2}{\Dl32}\right)\,.
\end{aligned}
\label{eq:l2check}
\end{equation}
We verified that the eigenvalues satisfy the characteristic equation to second order, see appendix \ref{ssec:characteristic
equation}.
The eigenvalues are correct at zeroth order to a fractional precision of about $10^{-4}$ or better, and through second order to
a precision of $10^{-8}$ or better.
In fact, the precision of $\lambda_1+\lambda_1^{(1)}+\lambda_1^{(2)}$ for $\sign(\Dmsqren)Y_e\rho E<0$ is completely saturated by the
limits of double precision computer calculations.

\subsection{Corrections to the eigenvectors}
Here we present the corrections to the eigenvectors which allows us to calculate the transition probabilities to
arbitrary order.
This was called the $V$-matrix approach in \cite{Minakata:1998bf}.

First, we relate the flavor eigenvectors to the zeroth order eigenvectors (no subscript) using $\U^m$, as in eq.~\ref{eq:MNS},
\begin{equation}
\begin{pmatrix}
\nu_e\\\nu_\mu\\\nu_\tau
\end{pmatrix}=\U^m
\begin{pmatrix}
\check\nu_1\\\check\nu_2\\\check\nu_3
\end{pmatrix}\,.
\end{equation}
Next, the exact eigenvectors of $\check{H}$, labeled with subscript (ex), are related to the eigenvectors of $\check{H}_0$ (the
zeroth order eigenvectors) by a
unitary matrix, which we call $W^\dagger$,
\begin{equation}
\begin{pmatrix}
\check\nu_1\\\check\nu_2\\\check\nu_3
\end{pmatrix}_{\negthickspace{\rm (ex)}}= W^\dagger
\begin{pmatrix}
\check\nu_1\\\check\nu_2\\\check\nu_3
\end{pmatrix}\,.
\label{eq:W}
\end{equation}
Combining the above gives,
\begin{equation}
\begin{pmatrix}
\nu_e\\\nu_\mu\\\nu_\tau
\end{pmatrix}=V
\begin{pmatrix}
\check\nu_1\\\check\nu_2\\\check\nu_3
\end{pmatrix}_{\negthickspace{\rm (ex)}}\, \quad 
{\rm where}  \quad V \equiv  \U^m W\,.
\end{equation}
The exact $V$ matrix transforms the exact eigenvectors of $\check H$  to the flavor basis.
In vacuum ($a=0$), $\U^m=\U$ and $W=\mathbbm1$, so $V=\U$ as expected.

Standard perturbation theory in $\check H_1$, which contains the small parameter $\epsilon'$, can be used to calculate $W^\dagger$.
Here we use a slightly modified perturbation theory to calculate $W$ directly.
Expanding $W$ as a power series in $\epsilon'$, we define
\begin{equation}
W\equiv W_0+ W_1+ W_2+\mathcal O(\eps'^3)\,.
\end{equation}
It is clear from eq.~\ref{eq:W} that $W_0=\mathbbm1$.

The first order correction to the $W$ matrix is given by
\begin{gather}
\begin{aligned}
(W_1)_{ij}&=
\begin{dcases}
0&i=j\\
-\frac{2E(\check H_1)_{ij}}{\Dl ij}\quad&i\neq j
\end{dcases}\,,\quad{\rm thus}\\[2mm]
W_1&=\eps'\Dmsqren
\begin{pmatrix}
&&-\frac{s_\psi}{\Dl31}\\[2mm]
&&\frac{c_\psi}{\Dl32}\\[2mm]
\frac{s_\psi}{\Dl31}\quad&-\frac{c_\psi}{\Dl32}
\end{pmatrix}\,.
\end{aligned}
\label{eq:W1}
\end{gather}

The second order correction, after using the facts that $\check H_1$ is symmetric and has no diagonal elements, eq.~\ref{eq:Hcheck1},
is 
\begin{gather}
\begin{aligned}
(W_2)_{ij} & =
\begin{dcases}
-\frac{1}{2}\sum_{k\neq i}\frac{[2E(\check H_1)_{ik}]^2}{(\Dl ik)^2}&i=j\\[2mm]
\frac{1}{\Dl ij} \sum_{k\neq i,k\neq j} \frac{2E(\check H_1)_{ik}~2E(\check H_1)_{kj}}{\Dl kj}&i\neq j
\end{dcases}\,,\quad{\rm thus}\\[4mm]
W_2 & = -\eps'^2\frac{(\Dmsqren)^2}2
\begin{pmatrix}
\frac{s_\psi^2}{(\Dl31)^2}&
-\frac{s_{2\psi}}{\Dl32\Dl21}\\[2mm]
\frac{s_{2\psi}}{\Dl31\Dl21}&
\frac{c_\psi^2}{(\Dl32)^2}\\
&&\left[\frac{c_\psi^2}{(\Dl32)^2}+\frac{s_\psi^2}{(\Dl31)^2}\right]
\end{pmatrix}\,.
\end{aligned}
\label{eq:W2}
\end{gather}
This series can be continued to reach arbitrary precision.
However, we have found that second order provides more than sufficient precision.

In summary the matrix relating the zeroth order eigenvalues of $\check H_0$ to the flavor basis is given by
\begin{equation}
V=\U^m W=U_{23}(\theta_{23}, \delta) U_{13}(\phi) U_{12}(\psi) (\mathbbm1+ W_1+W_2 )\,,
\label{eq:VofWdag}
\end{equation}
to second order in $\eps'$.
Demonstration of the unitary nature of $V$, to the appropriate order, is given in appendix \ref{ssec:unitarity}. 
With the eigenvalues and eigenvectors determined to second order we can now calculate the neutrino oscillation probabilities.

\section{Oscillation probabilities}
\label{sec:oscillation probabilities}
In vacuum and in matter with constant density, it is well known that the neutrino oscillation probabilities for $\nu_\alpha \to
\nu_\beta$ for \emph{three-flavor} mixing ($i, j = 1, 2, 3$) can be written in the following form\footnote{The equivalence of the
V-matrix method and the S-matrix method for calculating the oscillation probabilities is addressed in appendix \ref{ssec:V vs S}.}
\begin{gather}
\begin{aligned}
P(\nu_\alpha\to \nu_\beta) & = \left|\sum_{i=1}^{3} V^*_{\alpha i} V_{\beta i} e^{-i\frac{\lambda_i^{{\rm (ex)}}L}{2E}}\right|^2\\[1mm]
&\!
\begin{multlined}
=\delta^{\alpha\beta}+4C^{\alpha\beta}_{21}\sin^2\Delta_{21}+4C^{\alpha\beta}_{31}\sin^2\Delta_{31}+4C^{\alpha\beta}_{32}\sin^2\Delta_{
32}\\
+8D^{\alpha\beta}\sin\Delta_{21}\sin\Delta_{31}\sin\Delta_{32}\,,
\end{multlined}
\end{aligned}
\label{eq:general P}
\end{gather}
where
\begin{gather}
\begin{aligned}
C_{ij}^{\alpha\beta}&=-\Re[V_{\alpha i}V_{\beta i}^*V_{\alpha j}^*V_{\beta j}]\,,\\
D^{\alpha\beta}&=\phantom{-}\Im[V_{\alpha1}V_{\beta1}^*V_{\alpha2}^*V_{\beta2}]\,,\\
\Delta_{ij} & \equiv \Dl ij^{{\rm (ex)}}L/4E\, ,
\end{aligned}
\end{gather}
using the exact mixing matrix, $V_{\alpha i}$, and difference of the exact eigenvalues $\lambda_i^{{\rm (ex)}}$.
Both $V$ and $\lambda_i^{{\rm (ex)}}$s depend on the energy of the neutrino $E$, and the matter density $\rho$ but the baseline $L$,
dependence only appears in $\Delta_{ij}$. 

By unitarity
\begin{equation}
\sum_\beta P(\nu_\alpha\to\nu_\beta)=1\,,
\end{equation}
and using the fact that the $\sin^2$ functions and the triple sine function are linearly independent functions of L, as determined by
their non-zero
Wronskian, we have to the following powerful statements,
\begin{equation}
\sum_\beta C^{\alpha\beta}_{ij}=0\,,\quad
\sum_\beta D^{\alpha\beta}=0\,.
\end{equation}
Since $D^{\alpha\alpha}=0$, we also note that $D^{\alpha\beta}=-D^{\alpha\gamma}$ for $\alpha,\beta,\gamma$ all different.
So, up to one overall sign, there is only one $D$ term for all channels.

To determine the oscillation probability to n-th order in our perturbative expansion we must evaluate $C$, $D$, and $ \Dl ij^{{\rm
(ex)}} $
to the n-th order.
We denote this perturbative expansion as follows
\begin{gather}
\begin{aligned}
\Dl ij^{{\rm (ex)}}&=\Dl ij+\Dl ij^{(1)}+\Dl ij^{(2)}+\dots \\
C^{\alpha\beta}_{ij}&=(C^{\alpha\beta}_{ij})^{(0)}+(C^{\alpha\beta}_{ij})^{(1)}+(C^{\alpha\beta}_{ij})^{(2)}+\dots\\
D^{\alpha\beta}&=(D^{\alpha\beta})^{(0)}+(D^{\alpha\beta})^{(1)}+(D^{\alpha\beta})^{(2)}+\dots\,.
\end{aligned}
\end{gather}

\subsection{The zeroth order probabilities}
At zeroth order the $\Delta\lambda$'s are given by eq.~\ref{eq:lambdacheck} and the $C, D$ coefficients are the same as in vacuum with
$\theta_{13},\theta_{12}$ replaced with $\phi,\psi$ respectively, see eq.~\ref{eq:VofWdag}.
Therefore
\begin{equation}
\begin{aligned}
(C_{ij}^{\alpha\beta})^{(0)}&=-\Re[U_{\alpha i}U_{\beta i}^*U_{\alpha j}^*U_{\beta j}]\,,\\
(D^{\alpha\beta})^{(0)}&=\phantom{-}\Im[U_{\alpha1}U_{\beta1}^*U_{\alpha2}^*U_{\beta2}]\, ,
\end{aligned}
\label{eq:C0D0}
\end{equation}
where here the $U_{\alpha i}$ are elements of $\U^m=U_{23}(\theta_{23},\delta) U_{13}(\phi) U_{12}(\psi)$.
In table \subref{tab:zeroth} we give the zeroth order coefficients for $P(\nu_e\to \nu_e)$, $P(\nu_\mu\to\nu_e)$, and
$P(\nu_\mu\to\nu_\mu)$, from which all remaining transitions can be easily determined by unitarity.\footnote{The $\nu_\tau$
channels can also be obtained from the corresponding $\nu_\mu$ channel by the following replacements $c_{23} \to -s_{23}$ and  $s_{23}
\to c_{23}$.}

{\def\arraystretch{1.4}
\begin{sidewaystable}
\begin{subtable}{\linewidth}
\centering
\begin{tabular}{|l|c|c|c|}
\hline
$\nu_\alpha \to\nu_\beta$&
$(C_{31}^{\alpha \beta})^{(0)}$&
$(C_{21}^{\alpha\beta})^{(0)}$&
$(D^{\alpha\beta})^{(0)}$
\\\hline
$\nu_e\to\nu_e$&
$-c^2_\phi s^2_\phi c^2_\psi$&
$-c^4_\phi s^2_\psi c^2_\psi$&
0
\\\hline
$\nu_\mu\to\nu_e$&
$ s^2_\phi c^2_\phi c^2_\psi s^2_{23}+J^m_r \cos\delta $&
$c^2_\phi s^2_\psi c^2_\psi(c^2_{23}-s^2_\phi s^2_{23})+c_{2\psi}J^m_r\cos\delta$&
$-J^m_r\sin\delta$
\\\hline
\multirow{2}{*}{$\nu_\mu\to\nu_\mu$}&
$-c^2_\phi s^2_{23} (c^2_{23}s^2_\psi+s^2_{23}s^2_\phi c^2_\psi)$&
$-(c^2_{23}c^2_\psi+s^2_{23}s^2_\phi s^2_\psi)(c^2_{23} s^2_\psi+s^2_{23}s^2_\phi c^2_\psi)$&
\multirow{2}{*}{0}
\\
&
$-2 s^2_{23} J^m_r \cos\delta $&
$-2(c^2_{23}-s^2_\phi s^2_{23})c_{2\psi}J^m_{rr} \cos\delta+(2J^m_{rr}\cos \delta)^2 $&
\\\hline
\end{tabular}
\caption{The zeroth order coefficients for $C_{ij}^{\alpha\beta}$ and $D^{\alpha\beta}$ using eq.~\ref{eq:C0D0}.
The angles in matter, $\phi,\psi$, are given in sections \ref{ssec:U13} and \ref{ssec:U12}.
We also define the singly and doubly reduced Jarlskog coefficients in matter as $J^m_{r} \equiv  s_\psi c_\psi s_\phi c^2_\phi s_{23}
c_{23} $ and $J^m_{rr} \equiv J^m_{r}/ c^2_\phi $ respectively.
$(C_{32}^{\alpha\beta})^{(0)}$ can be obtained from $(C_{31}^{\alpha\beta})^{(0)}$ by using the $\lambda_{1,2}-\psi$ interchange
symmetry (eq.~\ref{eq:lambda12psi}) i.e.~$\lambda_1\leftrightarrow\lambda_2$, $c^2_\psi\leftrightarrow s^2_\psi$ and
$s_\psi c_\psi\to-s_\psi c_\psi$, which also changes the sign on the $J^m$'s.}
\label{tab:zeroth}
\end{subtable}
\\[12mm]
\stepcounter{table}
\begin{subtable}{\linewidth}
\centering
\begin{tabular}{|l|c|c|c|}
\hline
$\nu_\alpha\to\nu_\beta$&
$F^{\alpha\beta}_1$&
$G^{\alpha\beta}_1$&
$K^{\alpha\beta}_1$
\\\hline
$\nu_e\to\nu_e$&
$-2c^3_\phi s_\phi s^3_\psi c_\psi$&
$2s_\phi c_\phi s_\psi c_\psi c_{2\phi}$&
0
\\\hline
\multirow{2}{*}{$\nu_\mu\to\nu_e$}&
$c_\phi s^2_\psi[s_\phi s_\psi c_\psi(c^2_{23}+c_{2\phi}s^2_{23})$&
\multirow{2}{*}{$-2s_\phi c_\phi s_\psi(s^2_{23}c_{2\phi}c_\psi-s_{23}c_{23}s_\phi s_\psi\cos\delta)$}&
\multirow{2}{*}{$-s_{23}c_{23}c_\phi s^2_\psi(c^2_\phi c^2_\psi-s^2_\phi)\sin\delta$}
\\
&
$-s_{23}c_{23}(s^2_\phi s^2_\psi+c_{2\phi}c^2_\psi)\cos\delta]$
&
&
\\\hline
\multirow{2}{*}{$\nu_\mu\to\nu_\mu$}&
$2c_\phi s_\psi(s^2_{23}s_\phi c_\psi+s_{23}c_{23}s_\psi\cos\delta)\times$&
$-2c_\phi s_\psi(s^2_{23}s_\phi c_\psi+s_{23}c_{23}s_\psi\cos\delta)$&
\multirow{2}{*}{0}
\\
&
$(c^2_{23}c^2_\psi-2s_{23}c_{23}s_\phi s_\psi c_\psi\cos\delta+s^2_{23}s^2_\phi s^2_\psi)$&
$\times(1-2c^2_\phi s^2_{23})$&
\\
\hline
\end{tabular}
\caption{The functions $F_1^{\alpha\beta}$, $G_1^{\alpha\beta}$ and $K_1^{\alpha\beta}$, from eq.~\ref{eq:FGK}, are used to
calculate the first order coefficients $(C_{ij}^{\alpha\beta})^{(1)}$ and  $(D^{\alpha\beta})^{(1)}$ through eq.~\ref{eq:C1D1}.
$F_2^{\alpha\beta}$, $G_2^{\alpha\beta}$ and $K_2^{\alpha\beta}$ can be obtained using the $\lambda_{1,2}-\psi$ interchange symmetry
(eq.~\ref{eq:lambda12psi}) i.e.~$\lambda_1 \leftrightarrow \lambda_2$, $c^2_\psi \leftrightarrow s^2_\psi$ and $s_\psi
c_\psi\to-s_\psi c_\psi$.
The angles in matter, $\phi,\psi$, are given in sections \ref{ssec:U13} and \ref{ssec:U12}.}
\label{tab:first}
\end{subtable}
\end{sidewaystable}}
\stepcounter{table}
\clearpage

\subsection{The first order probabilities}
At first order the $\Delta\lambda$'s are again given by eq.~\ref{eq:lambdacheck}, since $\lambda_i^{(1)}=0$, see eq.~\ref{eq:Hcheck1},
because the diagonal elements of $\check  H_1$ are zero.
The first order corrections to $C, D$ only have terms proportional to $\Dl31^{-1},~\Dl32^{-1}$.
This comes from the form of $W_1$, eq.~\ref{eq:W1}, which follows from the position of the non-zero elements in $\check  H_1$.
In fact, all of the coefficients can be written in the following general form,
\begin{equation}
\begin{aligned}
(C_{21}^{\alpha\beta})^{(1)}&=\eps'\Dmsqren\left(\frac{F_1^{\alpha\beta}}{\Dl31}+\frac{F_2^{\alpha\beta}}{\Dl32}\right)\,,\\
(C_{31}^{\alpha\beta})^{(1)}&=\eps'\Dmsqren\left(\frac{F_1^{\alpha\beta}+G_1^{\alpha\beta}}{\Dl31}-\frac{F_2^{\alpha\beta}}{\Dl32}
\right)\,,\\
(C_{32}^{\alpha\beta})^{(1)}&=\eps'\Dmsqren\left(-\frac{F_1^{\alpha\beta}}{\Dl31}+\frac{F_2^{\alpha\beta}+G_2^{\alpha\beta}}{\Dl32}
\right)\,, \\
(D^{\alpha\beta})^{(1)}&=\eps'\Dmsqren\left(\frac{K_1^{\alpha\beta}}{\Dl31}-\frac{K_2^{\alpha\beta}}{\Dl32}
\right)\,,
\end{aligned}
\label{eq:C1D1}
\end{equation}
where the $F_{1,2}$, $G_{1,2}$ and $K_{1,2}$  are related by $\lambda_{1,2},\psi$ interchange previously discussed.
Thus only three modest expressions are required to describe the $C$'s and $D$ coefficients to first order for each channel.
The $F,G,K$ terms can be calculated from $\U^m$ by
\begin{equation}
\begin{aligned}
F_1^{\alpha\beta}&=-s_\psi\Re\left[(U_{\alpha1}U_{\beta3}^*
+U_{\alpha3}U_{\beta1}^*)U_{\alpha2}^*U_{\beta2}\right]\,,\\
G_1^{\alpha\beta}&=-s_\psi\Re\left[\left(U_{\alpha1}U_{\beta3}^*+U_{\alpha3}U_{\beta1}^*\right)
\left(2U_{\alpha3}^*U_{\beta3}-\delta_{\alpha\beta}\right)\right]\,. \\
K_1^{\alpha\beta} & =-s_\psi {\cal I} \left[(U_{\alpha 1}U^*_{\beta 3} + U_{\alpha 3}U^*_{\beta 1} ) U^*_{\alpha 2} U_{\beta 2}
\right]\,.
\end{aligned}
\label{eq:FGK}
\end{equation}
$F$ and $G$ are even under the interchange of $\alpha$ and $\beta$ whereas $K$ is
odd. Their explicit values are given in table \subref{tab:first}.

In the appearance channels the $CP$ violating term must be of the following form
\begin{equation}
D=\pm s_{12}c_{12} s_{13} c^2_{13} s_{23} c_{23} \sin_\delta
\frac{\prod_{i>j}\Delta m^2_{ij}}{\prod_{i>j}\Dl ij^{{\rm (ex)}}}\,,
\label{eq:D}
\end{equation}
where in the denominator one needs the exact eigenvalues in matter.
This is the Naumov-Harrison-Scott identity, see refs.~\cite{Naumov:1991ju,Harrison:1999df}.
We have checked this identity to the appropriate order, see appendix \ref{ssec:CPV}.

The $P(\nu_\alpha\to\beta)$ and $P(\bar\nu_\alpha\to\bar\nu_\beta)$ probabilities are related by $\delta\to-\delta$ and the
$P(\nu_\alpha\to\nu_\beta)$ and $P(\nu_\beta\to\nu_\alpha)$ transition probabilities are related by $L\to-L$.
From eq.~\ref{eq:general P}, we see that the $D$ term is the only term odd in $L$.
From tables \subref{tab:zeroth} and \subref{tab:first}, we see that the $D$ term is also the only one odd in $\delta$, confirming the
CPT invariance of these equations.
Moreover, all of the $D^{\alpha\beta}$ terms are the same order by order up to a coefficient of $-1,0,1$.

\subsection{The second order probabilities}
Although we have not expanded the second order oscillation probabilities analytically, the second order corrections to the
eigenvalues, $\lambda_i^{(2)}$, as well as the second order corrections to the mixing matrix, $W_2$, have been used to calculate
the oscillation probabilities to second order. The resulting oscillation probabilities are more than two orders of magnitude closer
to the exact values than the first order probabilities.

\subsection{Precision of the perturbation expansion}
\label{ssec:precision}
\begin{figure}
\centering
\includegraphics[width=0.75\textwidth]{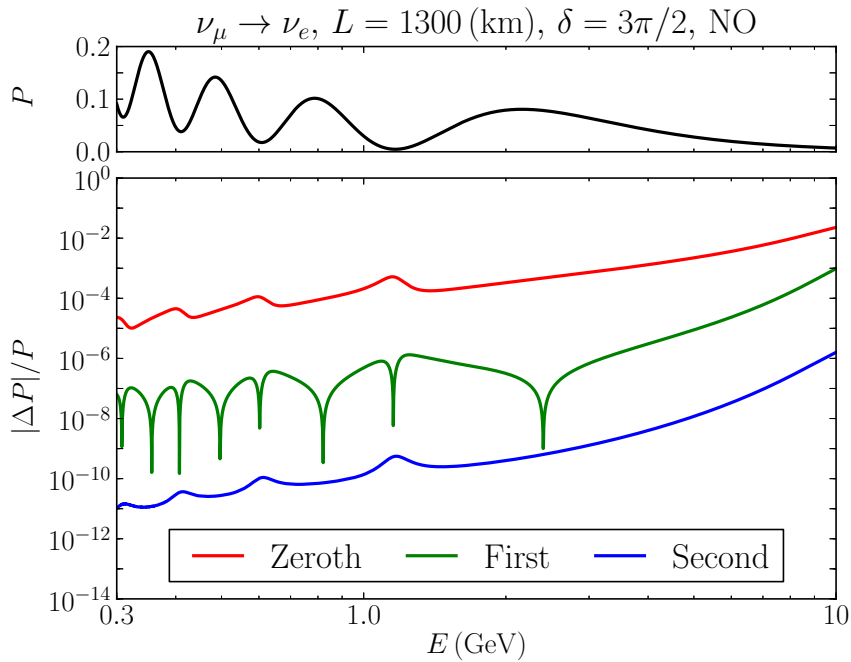}
\caption{The $\nu_\mu\to\nu_e$ oscillation probability is plotted in the upper part of the figure for DUNE parameters; a 1300 km
baseline and $Y_e\rho=1.4$ g$\cdot$cm$^{-3}$.
The fractional uncertainties at zeroth and first order are plotted using the analytic formulas in tables \subref{tab:zeroth} and
\subref{tab:first} respectively.
The probability to second order is calculated by using $\lambda$'s and $W$ through second order, see eqs.~\ref{eq:l2check} and
\ref{eq:W2}}.
\label{fig:Pmu2e_Precision}
\end{figure}

The oscillation probabilities that were perturbatively calculated in this section are only useful if they are more precise than the
experimental uncertainties.
In figure \ref{fig:Pmu2e_Precision}, we have plotted the fractional uncertainties\footnote{The exact oscillation probability were
calculated using \cite{Zaglauer:1988gz,Kimura:2002wd}.} at each order of our perturbative expansion for the 
$\nu_\mu \to \nu_e$ channel at the DUNE \cite{Acciarri:2015uup}, baseline of 1300~km. 
The precision at the first oscillation maximum and minimum for DUNE are shown in table \ref{tab:accuracy}.
We note that the precision improves at lower energies, such as for NO$\nu$A \cite{Patterson:2012zs} and T2K/T2HK \cite{Abe:2011ks,
Abe:2015zbg}.
The results are comparable for different values of $\delta$, for the inverted ordering, for other channels, and for antineutrino
mode. 
Therefore, even at zeroth order, the precision exceeds the precision of the expected experimental results.

\def\arraystretch{1.2}
\begin{table}
\centering
\begin{tabular}{|c|c|c|c|}
\hline
\multicolumn{2}{|c|}{DUNE: NO, $\delta=3\pi/2$}&First min&First max\\\hline
\multicolumn{2}{|c|}{$P(\nu_\mu\to\nu_e)$}&0.0047&0.081\\\hline
\multicolumn{2}{|c|}{$E$ (GeV)}&1.2&2.2\\\hline
\multirow{3}{*}{\quad$\dfrac{|\Delta P|}P$\quad}
&Zeroth&$5\e{-4}$&$4\e{-4}$\\\cline{2-4}
&First&$3\e{-7}$&$2\e{-7}$\\\cline{2-4}
&Second&$6\e{-10}$&$5\e{-10}$\\\hline
\end{tabular}
\caption{The transition probabilities, energies, and fractional uncertainties at zeroth, first, and second order.
Values are calculated at DUNE for $\nu_\mu\to\nu_e$ with the NO and $\delta=3\pi/2$.
At higher maxima and minima the fractional uncertainties are even smaller.}
\label{tab:accuracy}
\end{table}

\section{Conclusions}
\label{sec:conclusion}
In this paper we have further developed and expanded upon the recent perturbative framework for neutrino oscillations in uniform
matter, introduced in \cite{Minakata:2015gra}.   The new oscillation probabilities are of the same simple, compact functional form with
slightly more complicated coefficients, yet, the range of applicability now includes the whole $L/E$ versus matter potential, $a$,
plane, i.e.~the restriction that $L/E$ be small, ($L/E \ll 1/\Delta m^2_{21}$) around the vacuum values of the matter potential has
been completely removed. In fact, with these new improvements, the oscillation probabilities in vacuum are exact at zeroth order in our
perturbative expansion.
This occurs because the expansion parameter   $ s_{12} c_{12} \Delta m^2_{21}/\Dmsqren=0.014$ is further multiplied by
$\s$, where $\phi$ is the mixing angle $\theta_{13}$ in matter.  In vacuum, $\phi=\theta_{13}$ and therefore all
corrections to zeroth order vanish. 

To achieve this extended range of applicability, an additional rotation of the Hamiltonian is performed over that in
\cite{Minakata:2015gra}.  The third angle $\psi$ is the mixing angle $\theta_{12}$ in matter.  In the resulting
Hamiltonian, the diagonal elements are the eigenvalues of the zeroth order Hamiltonian and do not cross for any values of the matter
potential, especially near the solar resonance (this occurred in \cite{Minakata:2015gra}).  The non-diagonal elements of the new
Hamiltonian are the perturbing Hamiltonian for our perturbative expansion and their size is controlled by the small parameter    $ \s 
s_{12} c_{12} \Delta m^2_{21}/\Dmsqren$, mentioned in the previous paragraph.
The new perturbative expansion is now well defined  for
all values of the matter potential and gives very accurate oscillation probabilities. 
We have performed many cross checks on the perturbative expansion, e.g.~we have
checked the $CP$ violating term recovers, order by order,  the known form.
We have calculated the oscillation probabilities
for zeroth, first, and second order in our expansion parameter.   
For most practical applications related to experiments, the zeroth
order oscillation probabilities are  sufficiently accurate with a typical fractional uncertainty of better than $10^{-3}$. 
Including the first and second order corrections the accuracy improves that to better than $10^{-6}$ and $10^{-9}$, respectively.

\acknowledgments
P.B.D.~acknowledges support from the Fermilab Graduate Student Research Program in Theoretical Physics operated by Fermi Research
Alliance, LLC. This work is also supported in part by DOE grant DE-SC0011981.

H.M.~thanks Instituto de F\'{\i}sica, Universidade de S\~ao Paulo for the great opportunity of stay under support by Funda\c{c}\~ao de
Amparo \`a Pesquisa do Estado de S\~ao Paulo (FAPESP) with grant number 2015/05208-4.
He thanks Fermilab Theory Group for warm hospitality in his visits.

S.P.~acknowledges partial support from the  European Union FP7  ITN INVISIBLES (Marie Curie Actions, PITN-GA-2011-289442).
This project has received funding from the European Union's Horizon 2020 research and innovation programme under the Marie
Sk\l{}odowska-Curie  grant agreement No 690575-InvisiblesPlus RISE. This project has received funding from the European Union's
Horizon 2020 research and innovation programme under the Marie Sk\l{}odowska-Curie grant agreement No 674896-Elusives ITN.  

Fermilab is operated by the Fermi Research Alliance, LLC  under contract no.~DE-AC02-07CH11359 with the U.S.~Department of Energy.

\appendix
\section{Technical details}
\label{sec:technical details}

\subsection{Generalized approach to diagonalization}
\label{ssec:general diagonalization}
We describe the diagonalization of a particular $2\times2$ submatrix and the angle and eigenvalues.
This is the approach used twice in subsections \ref{ssec:U13} and \ref{ssec:U12} to diagonalize the 1-3 and then the 1-2 submatrices.

Given a general symmetric $2\times2$ matrix we wish to diagonalize with angle $\phi$, we write
\begin{equation}
\begin{pmatrix}
\lambda_\sigma\\&\lambda_\rho
\end{pmatrix}
=U(\phi)^\dagger
\begin{pmatrix}
\lambda_a&\lambda_x\\
\lambda_x&\lambda_c
\end{pmatrix}U(\phi)\,,
\label{eq:general:diagonalization}
\end{equation}
where
\begin{equation}
U(\phi)\equiv
\begin{pmatrix}
c_\phi&s_\phi\\
-s_\phi&c_\phi
\end{pmatrix}\,.
\end{equation}
Since trace and determinant are unchanged by the $U$ sandwich,
\begin{equation}
\lambda_\sigma+\lambda_\rho=\lambda_a+\lambda_c\quad{\rm and}\quad
\lambda_\rho\lambda_\sigma=\lambda_a\lambda_c-\lambda^2_x\,.
\label{eq:trace and determinant identities}
\end{equation}
By squaring the trace equation and subtracting 4 times the determinant equation we have 
\begin{equation}
(\lambda_\rho-\lambda_\sigma)^2=(\lambda_a-\lambda_c)^2+4 \lambda^2_x\,,
\end{equation}
thus
\begin{equation}
\lambda_{\rho,\sigma}=\frac12\left[\left(\lambda_a+\lambda_c\right)\pm\sqrt{\left(\lambda_a-\lambda_c\right)^2+4\lambda_x^2}\right]\,.
\end{equation}
Next, we rewrite eq.~\ref{eq:general:diagonalization} by left (right) multiplying by $U(\phi)$ ($U^\dagger(\phi)$), then
\begin{equation}
U(\phi)
\begin{pmatrix}
\lambda_\sigma\\&\lambda_\rho
\end{pmatrix}
U(\phi)^\dagger=
\begin{pmatrix}
c^2_\phi\lambda_\sigma+s^2_\phi\lambda_\rho&s_\phi c_\phi(\lambda_\rho-\lambda_\sigma)\\
s_\phi c_\phi(\lambda_\rho-\lambda_\sigma)&s^2_\phi\lambda_\sigma+c^2_\phi\lambda_\rho 
\end{pmatrix}=
\begin{pmatrix}
\lambda_a&\lambda_x\\
\lambda_x&\lambda_c
\end{pmatrix}\,.
\end{equation}
This gives us three equations,
\begin{equation}
\begin{aligned}
\lambda_a&=c^2_\phi\lambda_\sigma+s^2_\phi\lambda_\rho\,,\\
\lambda_c&=s^2_\phi\lambda_\sigma+c^2_\phi\lambda_\rho\,,\\
\lambda_x&=(\lambda_\rho-\lambda_\sigma)s_\phi c_\phi\,.
\label{eq:lambda mixing}
\end{aligned}
\end{equation}
The last equation is the standard equation for $s_{2\phi}$.
Subtracting (adding) the first two gives the standard equation for $c_{2\phi}$ (the trace).
Thus the rotation angle is defined by the following
\begin{equation}
\lambda_x=(\lambda_\rho-\lambda_\sigma)s_\phi c_\phi\quad{\rm and}\quad
(\lambda_c-\lambda_a)=(\lambda_\rho-\lambda_\sigma)(c^2_\phi-s^2_\phi)\,.
\end{equation}
In addition, using only $c_\phi^2+s_\phi^2=1$ we can write down the following useful identities
\begin{equation}
\begin{aligned}
c^2_\phi&
=\frac{\lambda_\rho-\lambda_a}{\lambda_\rho-\lambda_\sigma}
=\frac{\lambda_c-\lambda_\sigma}{\lambda_\rho-\lambda_\sigma}\,,\\
s^2_\phi&
=\frac{\lambda_\rho-\lambda_c}{\lambda_\rho-\lambda_\sigma}
=\frac{\lambda_a-\lambda_\sigma}{\lambda_\rho-\lambda_\sigma}\,,
\label{eq:angle squared}
\end{aligned}
\end{equation}
which are used extensively throughout this paper.
This set of operations will be used both for $\phi$ and $\psi$ rotations.

\subsection{Useful identities}
\label{ssec:useful identities}
From the trace and determinant identities, see eq.~\ref{eq:trace and determinant identities},
\begin{align}
\lambda_-+\lambda_+&=\lambda_a+\lambda_c\,,\\
\lambda_1+\lambda_2&=\lambda_-+\lambda_0\,,
\end{align}
\begin{align}
\lambda_+\lambda_-&=\lambda_a\lambda_c-\left[\Dmsqren c_{13}s_{13}\right]^2\,,\\
\lambda_1\lambda_2&=\lambda_0\lambda_--\left[\eps\Dmsqren c_{12}s_{12}\cc\right]^2\,,
\end{align}
where we recall that the $\lambda_{a,b,c}$ in the tilde basis are defined in eq.~\ref{eq:lambda abc}.
Another useful relation is
\begin{equation}
\cc\s=s_{13}c_{13}\frac a{\Dl+-}\,,
\end{equation}
then for $a\ll\Dmsqren$,
\begin{equation}
\s\approx s_{13}c_{13}\frac a{\Dmsqren}\,.
\end{equation}

\subsection{Limits}
\label{ssec:limits}
We list the values of the angles and the eigenvalues in vacuum and for $a\to\pm\infty$ in table~\ref{tab:limits}.
{\def\arraystretch{1.2}
\begin{table}
\centering
\begin{tabular}{|c|c|c|c|}
\hline
$a$&0&$-\infty$&$+\infty$\\\hline
$\phi$&$\theta_{13}$&$0\,(\pi/2)$&$\pi/2\,(0)$\\
$\psi$&$\theta_{12}$&0&$\pi/2$\\\hline
$\lambda_-$&$s_{12}^2\Delta m^2_{21}$
&$\lambda_a\,(\lambda_c)$
&$\lambda_c\,(\lambda_a)$\\
$\lambda_0$&$c_{12}^2\Delta m^2_{21}$&$\lambda_b$&$\lambda_b$\\
$\lambda_+$&$\Delta m^2_{31}$&
$\lambda_c\,(\lambda_a)$&
$\lambda_a\,(\lambda_c)$\\\hline
$\lambda_1$&0&$\lambda_a\,(\lambda_c)$&$\lambda_b$\\
$\lambda_2$&$\Delta m^2_{21}$&$\lambda_b$&$\lambda_c\,(\lambda_a)$\\
$\lambda_3$&$\Delta m^2_{31}$&
$\lambda_c\,(\lambda_a)$&
$\lambda_a\,(\lambda_c)$\\\hline
\end{tabular}
\caption{The NO (IO) limits of the angles and the eigenvalues in vacuum and for $a\to\pm\infty$,
where $\lambda_a=a+(s^2_{13}+\eps s^2_{12})\Dmsqren$, $\lambda_b=\eps c^2_{12}\Dmsqren$, and $\lambda_c=(c^2_{13}+\eps
s^2_{12})\Dmsqren$, from eq.~\ref{eq:lambda abc}.}
\label{tab:limits}
\end{table}}

\subsection{Characteristic equation}
\label{ssec:characteristic equation}
The characteristic equation for neutrino oscillation in matter is
\begin{multline}
\lambda^3-\left(\Delta m_{21}^2+\Delta m_{31}^2+a\right)\lambda^2+\left\{\Delta m_{21}^2\Delta
m_{31}^2+a\left[(c_{12}^2+s_{12}^2s_{13}^2)\Delta
m_{21}^2+c_{13}^2\Delta m_{31}^2\right]\right\}\lambda\\
-\left(ac_{12}^2c_{13}^2\Delta m_{21}^2\Delta m^2_{31}\right)=0\,.
\end{multline}
The coefficient of the $\lambda^2$ term is the sum of the eigenvalues, the coefficient of the $\lambda$ term is the sum of pairs of the
eigenvalues, and the coefficient of the $\lambda^0$ term is the triple product of eigenvalues.

We now verify that our matter mass eigenvalues satisfy these expressions to second order.
First, the $\lambda_{-,0,+}$ eigenvalues satisfy the first requirement exactly as was discussed in \cite{Minakata:2015gra}.
Since $\sum_{i=1,2,3}\lambda_i=\sum_{i=-,0,+}\lambda_i$, so the $\lambda_{1,2,3}$ eigenvalues also satisfy the first requirement.
Also, from eq.~\ref{eq:l2check}, $\sum_{i=1,2,3}\lambda_i^{(2)}=0$, so the $\lambda_{1,2,3}$ eigenvalues also satisfy the first
requirement exactly through second order.
We have also verified that each of the other two conditions are satisfied two second order.

\subsection{Unitarity of the \texorpdfstring{$W$}{W} matrix}
\label{ssec:unitarity}
We verify that the $V$ matrix satisfies the unitarity requirements, $VV^\dagger=\mathbbm1$.
$\U^m$ is unitary by definition.
Then we just need that the $W$ matrix is unitary.
The zeroth order requirement is $W_0W_0^\dagger=\mathbbm1$ which is immediately satisfied since $W_0=\mathbbm1$.
At first order the requirement is $W_1+W_1^\dagger=0$.
This is equivalent, to the requirement that $W_1$ is anti-Hermitian, or that $\check H_1$ is Hermitian, which they are, respectively,
see eq.~\ref{eq:W1}.

To second order, the unitarity requirement becomes, $W_2+W_2^\dagger=-W_1^2$.
That is, that the Hermitian part of $W_2$ must be $-W_1^2/2$, which it is.
An additional anti-Hermitian part is unconstrained and is calculated through perturbation theory.

\subsection{\texorpdfstring{$V$}{V}-matrix, \texorpdfstring{$S$}{S}-matrix comparison}
\label{ssec:V vs S}
In the S-matrix method, the oscillation probabilities are given by, see for example \cite{Minakata:2015gra},
\begin{gather}
\begin{aligned}
S_S(L)={}&\U^m~e^{-iH_0L}\Omega(L)~(\U^m)^\dagger\\
\Omega(L)={}&1+(-i)\int^L_0dx~e^{iH_0x}H_1e^{-iH_0x}\\
&+(-i)^2\int^L_0dx~e^{iH_0x}H_1e^{-iH_0x}\int^x_0 dx'~e^{iH_0x'}H_1e^{-iH_0x'}+\cdots\,.
\end{aligned}
\end{gather}
where $H_0$ and $H_1$ are given by eqs.~\ref{eq:Hcheck0} and \ref{eq:Hcheck1}.
(We drop the ``check'' in this appendix.)

In the V-matrix method, used in this paper, the oscillation probabilities are given by,
\begin{gather}
\begin{aligned}
S_V(L)  & = \U^m  ~W  e^{-i \Lambda L/2E}  W^\dagger ~(\U^m)^\dagger \\
(\Lambda)_{ij} & = \delta_{ij} (\lambda_i + \lambda^{(1)}_i + \lambda^{(2)}_i + \cdots )  \\
W & = 1+W_1+W_2 +\cdots\,,
\end{aligned}
\end{gather}
where the $\lambda_i /2E$ are the eigenvalues of $H_0$.
$\lambda^{(n)}_i$ and $W_n$ are given by n-th order perturbation theory.

Specializing to the case when the perturbing Hamiltonian has no diagonal elements,
\begin{equation}
(H_1)_{ij}=(1-\delta_{ij})h_{ij}/2E\,,
\end{equation}
which is relevant for the perturbation discussed in this paper, $W$ can be calculated from eq.~\ref{eq:W1} for first order and
eq.~\ref{eq:W2} for second order.

Then it is trivial to show that to first order,
\begin{gather}
\begin{aligned}
[(\U^m)^\dagger S_S(L)  \U^m  ]_{ij}  & =[  (\U^m)^\dagger S_V(L)  \U^m ]_{ij} \\
&=\delta_{ij}e^{-i\lambda_i L/2E}+(1-\delta_{ij}) ~\frac{h_{ij}}{\Dl ij} \left(e^{-i\lambda_i L/2E}- e^{-i\lambda_jL/2E}\right)\,.
\end{aligned}
\end{gather}
 We have also checked that they are equal at second order.
As this is just a consistency check of perturbation theory, we postulate that it is true to all orders, without presenting an all
orders proof.

\subsection{\texorpdfstring{$CP$}{CP} violating term}
\label{ssec:CPV}
It is useful to rewrite the numerator of eq.~\ref{eq:D} as $\eps(\Dmsqren)^3(1-\eps\cos2\theta_{12}-\eps^2c_{12}^2s_{12}^2)$.
We evaluate $D^{e\mu}$ through first order, keeping terms that are explicitly second order in $\eps$, noting that dividing by $\Dl21$
introduces an additional factor of $\eps$ in vacuum.
\begin{equation}
(D^{e\mu})^{(0)}+(D^{e\mu})^{(1)}=s_\delta J_r\frac{\eps(\Dmsqren)^3(1-\eps\cos2\theta_{12})}{\Dl21\Dl31\Dl32}\,,
\end{equation}
where $J_r$ is the reduced Jarlskog factor, see ref.~\cite{Jarlskog:1985ht},
\begin{equation}
J_r\equiv c_{12}s_{12}c_{13}^2s_{13}c_{23}s_{23}\,.
\end{equation}
The dropped higher order contribution to the numerator is
\begin{multline}
-\eps^2c_{12}^2s_{12}^2\frac{\Dl+-+(\Dmsqren-a)}{4(\Dl+-)^3}\times\\
\left[(\Dmsqren)^2+3a^2-4c_{2\theta_{13}}a\Dmsqren+(\Dmsqren+a)\Dl+-\right]\,,
\end{multline}
which is $-\eps^2c_{12}^2s_{12}^2$ in vacuum as desired since $\Dl+-$ is $\Dmsqren$ in vacuum.

\addcontentsline{toc}{section}{References}
\bibliographystyle{JHEP}
\bibliography{Neutrino_Perturbation}

\end{document}